\documentclass{aa}
\usepackage{graphicx}
\usepackage{natbib}
\usepackage{txfonts}
\usepackage{epstopdf}
\usepackage[utf8]{inputenc}
\usepackage[english]{babel}
\bibpunct{(}{)}{;}{a}{}{,} % to follow the A&A style
\usepackage{fontawesome}
\usepackage[usenames,dvipsnames]{color}
\usepackage[colorlinks=true,citecolor=blue,linkcolor=magenta,urlcolor=Cerulean]{hyperref}

\usepackage{caption}
\definecolor{twitterblue}{RGB}{64,153,255}
\newcommand{\twitter}[1]{\href{https://twitter.com/#1}{\textcolor{twitterblue}{\faTwitter}\,\tt\hspace{2pt}\textcolor{blue}{@#1}}}

\def\apjs{ApJS}

\def\aap{Astron. Astrophys.}

\begin{document}

\title{Inflection point in the power spectrum of stellar brightness variations III:\\ Facular vs. Spot dominance on stars with known rotation periods.}
\titlerunning{ --  GPS III: Faculae vs. Spot dominance on stars with known rotation periods.}

\author{E.M.~Amazo-G\'{o}mez \inst{1,2} 
\and A.I.~Shapiro \inst{1} 
\and S.K.~Solanki \inst{1,3}
\and G.~Kopp \inst{4} 
\and M.~Oshagh\inst{2,5}
\and \\ T.~Reinhold \inst{1}
\and~A.~Reiners\inst{2}}
\offprints{E.M.~Amazo-G\'{o}mez\\
\twitter{AstroSumerce}}

\institute{Max-Planck-Institut f\"{u}r Sonnensystemforschung, Justus-vonustus-von-Liebig-Weg 3, 37077 G\"{o}ttingen, Germany\\
\email{amazo@mps.mpg.de}
\and Georg-August Universit\"{a}t G\"{o}ttingen, Institut f\"{u}r Astrophysik, Friedrich-Hund-Platz 1, 37077 G\"{o}ttingen, Germany
\and School of Space Research, Kyung Hee University, 446-701 Yongin, Gyeonggi, Korea
\and Laboratory for Atmospheric and Space Physics, 3665 Discovery Dr., 80303 Boulder, CO, USA
%\and Station of Muggle Observation, Gryffindor College, Hogwarts, Magical forest, UK
\and Instituto de Astrof\'isica de Canarias (IAC), E-38200 La Laguna, Tenerife, Spain}
%\and Goofinov Institut, 10 Jokester Way, Punville, G\"{o}ttebekiddinme, Germany}
\authorrunning{E.~M.~Amazo-G\'omez et al.}

\date{Received ; accepted}

\abstract
% context heading (optional}
{Stellar rotation periods can be determined by observing brightness variations caused by active magnetic regions transiting visible stellar disk as the star rotates. 
The successful stellar photometric surveys stemming from the \textit{Kepler} and \textit{TESS} observations led to the determination of rotation periods in tens of thousands of young and active stars. However, there is still a lack of information about rotation periods of  older and less active stars, like the Sun. The irregular temporal profiles of light curves caused by the decay times of active regions, which are comparable to or even shorter than stellar rotation periods, combine with the random emergence of active regions to make period determination for such stars very difficult} 
{We tested the performance of the new method for the determination of stellar rotation periods against stars with previously
determined rotation periods. The method is based on calculating the gradient of the power spectrum (GPS) and identifying the position of the inflection point (i.e. point with the highest gradient). The GPS method is specifically aimed at determining rotation periods of low-activity stars like the Sun.}
{We applied the GPS~method to 1047~Sun-like stars observed by the \textit{Kepler} telescope. We separately considered two stellar samples: one with near-solar rotation periods (24--27.4 d) and broad range of effective temperatures (5000--6000 K), another with  near-solar effective temperatures (5700--5900 K) and broad range of rotation periods (15--40 d).}
{We show that the GPS~method returns precise values of stellar rotation periods. Furthermore, it allows us to constrain the ratio between facular and spot areas of active regions at the moment of their emergence. We show that relative facular area decreases with stellar rotation rate.}
%Conclusions
{Our results suggest that the GPS method can be successfully applied to retrieve periods of stars with both regular and non-regular light curves.}


\maketitle
\section{Introduction}\label{sec:intro}

\begin{figure*}[!ht]\label{fig1}
\includegraphics[trim={0 150 0 150 cm},clip,width=1.\textwidth]{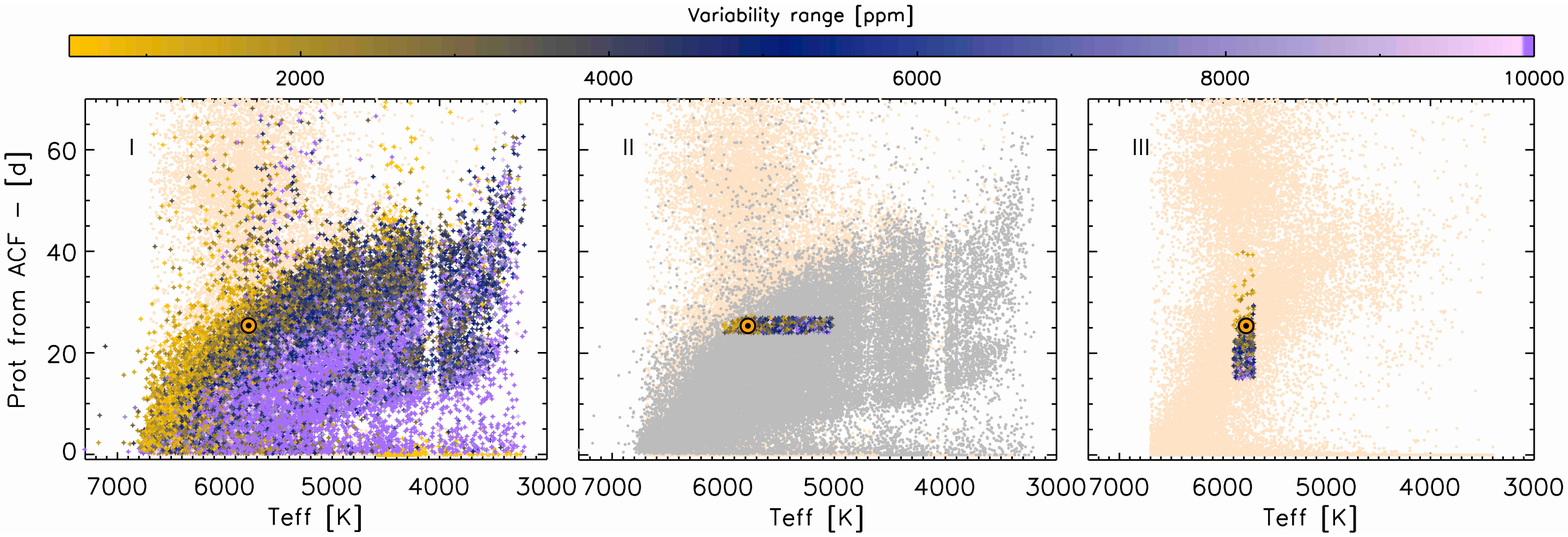}
 \caption{Panel I: temperature-rotation diagram for a sample of 34030 stars (coloured circles indicating the variability range) with rotation periods determined by \cite{2014ApJS..211...24M} and 55501 stars where they found a period but deemed it to be not significant (bisque dots, See panel-III for better visualisation).  Panel~II sample of 34030 stars with rotation periods determined coloured in grey and 55501 stars with not significant rotation period determination in bisque colour. For Panels~II and III only stars from sample~A (panel II) and sample B (panel III) are shown in colour, see Table~\ref{tab:1} for the properties of samples A and B. Panels~II illustrates the stellar sample A, selected by near solar rotation period and temperatures from 5000 K to 6000 K. Panel~III, illustrates stellar sample B, that contains stars with near solar effective temperature and a broad range in rotation periods. The Sun is represented by the solar symbol~$\odot$.} \label{Fig1_GPS_III}
\end{figure*}

Rotation periods in cool main-sequence stars can be traced by observing the brightness modulation caused by the presence of active regions on stellar surfaces. 
Those active regions are generated by the emergence of strong localised magnetic fields approximately described by flux tubes \citep[see, e.g.][]{solanki1993}. Large flux tubes form dark spots, while ensembles of smaller flux tubes form bright faculae \citep[see, e.g,][for a detailed review of the solar case]{2006RPPh...69..563S}. The active regions usually consist of a sunspot group surrounded by faculae. The transits of such active regions over the visible disk as the star rotates would cause brightness variability. Consequently, the stellar light curves (LCs) contain information about both, rotation periods and properties of active regions. However, retrieving this information from the light curves often appears to be a daunting task \citep[see, e.g.,][]{Basri2018}.

The \textit{Kepler} mission \citep{2010Sci...327..977B} has provided records of photometric observations with unprecedented precision and cadence. \textit{Kepler} light curves have been widely used to determine stellar rotation periods \citep[e.g.,][]{Walkowicz2013,2015A&A...583A..65R,2013A&A...557L..10N,Garcia2014,2014ApJS..211...24M,2016JSWSC...6A..38B,2018MNRAS.474.2094A,Santos_2019}. Despite the success in determining rotation periods of many fast rotating and active stars \citep[see, e.g.,][who published rotation periods of about 34030 stars identified as on the main sequence]{2014ApJS..211...24M} there is a lack of information on periods of slowly rotating stars, i.e. stars with near-solar and longer rotation periods. For example, the rotational period of the Sun may not be detectable during intermediate and high levels of solar activity~\citep[see,][]{2014MNRAS.443.1451L,2015MNRAS.450.3211A}. 

The difficulties in detecting periods of slowly rotating stars might be an important contributor to the explanation of lower-than-expected numbers of G-type stars with near-solar rotation periods \citep{2019ApJ...872..128V}.
The difficulty in reliably measuring rotation periods of stars with variability patterns similar to that of the Sun can also affect solar-stellar comparison studies \citep[see, e.g.][]{witzke2020,Reinhold518}.

In this context we have developed a method
aimed at determining rotation periods of low-activity stars like the Sun. In \citet{paperI} (hereinafter, Paper\,I) we found that the power spectra of brightness variations of such stars are strongly affected by the evolution of active regions. In particular, the rotation peak can be significantly weakened or even disappear from the power spectrum if the lifetimes of starspots are too short. Furthermore, the delicate balance between spot and facular contributions to the variability might lead to the appearance of spurious peaks, which do not correspond to the rotation period but could be easily mistaken for it \citep[see also][]{Shapiro2017}. 

\begin{table*}
\caption{Stellar parameters for samples~A~and~B.}
\begin{tabular}{l c c c c c c}
\hline
\hline
& & & & & & \\[-4pt]
Sample & N & $T_{\rm eff}^{(1)}$[K] & $\log\,{\rm g}^{(1)}$ & [Fe/H]$^{(1)}$ & Var$^{(2)}$[ppm] & $P_{\rm rot}^{(2)}$ [d] \\ 
& & & & & & \\[-1pt]
\hline
\hline
& & & & & & \\[-3pt]
%5015-5991
A & 686 & 5000-6000 & 4.20-4.69 & -1.46-0.56 & 211-39748 & 24.0-27.4 \\
%5701-5898
B & 361 & 5700-5900 & 4.21-4.60 & -1.08-0.44 & 211-17530 & 15.0-39.8 \\
Sun & 1 & 5778 & 4.44 & 0.0 & 300-1500 & 27.27 (Sy), 25.38 (Sid) \\
%C & 471 & 5303-6109 & 4.21-5.08 & -1.00-0.51 & 800-28300 & 14.7-36.1 \\
\hline
\end{tabular}
%\vspace{0.4cm}
%\captionsetup{labelformat=empty}
\centering
\tablefoot{Stellar parameters for stellar samples~A~and~B. 1) Effective temperature ($T_{\rm eff}$), surface gravity ($\log\, {\rm g}$), and metallicity ([Fe/H]) values are taken from \cite{Huber2014}. 2) Variability range (Var) and rotation periods ($P_{\rm rot}$) are taken from \cite{2014ApJS..211...24M}. We take the solar synodic (Sy) and sidereal (Sid) Carrington rotation period values as reference.}\label{tab:1} 
\end{table*}

In Paper\,I we showed that the high-frequency tail of the power spectrum is much less affected by the evolution of magnetic features than frequencies near the rotation period. Consequently, we proposed to use information in the high-frequency tail for the determination of stellar rotation periods. In particular, we suggested that the period $P_{\rm HFIP}$ corresponding to the maximum of the gradient of the power spectrum (GPS) (i.e. to the inflection point) in the high-frequency tail could be used to identify the stellar rotation period, $P_{\rm rot}$, via the simple scaling relation:

\begin{equation}~\label{eq1}
P_{\rm rot}\,=\,P_{\rm HFIP} / \alpha. 
\end{equation}

Here $\alpha$ is a calibration factor which is independent of the evolution of active regions. It shows only a very weak dependence on the stellar inclination. For example, the inclination dependence can be neglected for inclinations of 45$^{\circ}$ and greater, see Fig.~9 from Paper\,I. Statistically this corresponds to roughly 70\% of stars.

The model developed in Paper\,I indicated that the value of $\alpha$ shows a moderate dependence on the ratio between facular and spot areas of the {\it individual} active regions at the moment of emergence, $S_{\rm fac}/S_{\rm spot}$. This ratio was assumed to be the same for all active regions (see a detailed discussion in Paper\,I). The dependence of the inflection point position on the facular-to-spot area ratio leads to a certain degree of uncertainty (up to 25\%) in determining stellar rotation periods since the value of $S_{\rm fac}/S_{\rm spot}$ for a given star is {\it a prior} unknown. At the same time, it allows retrieving valuable information about facular- vs. spot-dominated regimes of the variability for stars where rotation periods can be determined using other methods \cite[see,][]{Eliana1} (hereinafter, Paper\,II). 

A first test of the gradient of the power spectrum method (hereinafter, the GPS method) has been performed in Paper\,II where we have applied it to solar brightness variations. We showed that, in contrast to other methods, GPS allows an accurate determination of the solar rotation period at all levels of solar activity. Additionally, we analysed time intervals when solar variability was spot-dominated and when it was faculae-dominated. We showed that these regimes can be distinguished in the GPS profile due to the substantially different center-to-limb variations of faculae and spots.

In this study, we apply the GPS method to stars with determined rotation periods from \textit{Kepler} photometry. The goal is twofold: First, we further test the GPS method before applying it to stars with unknown rotation periods; and second, we investigate whether there is a dependence of the $\alpha$\,factor, and consequently the facular/spot composition of stellar active regions, on the rotation period. In section\,\ref{sec:2}, we describe the stellar sample used. In section\,\ref{sec:3}, we present the main results, while the Conclusions are summarized in section\,\ref{sec:4}.

\section{Stellar sample selection}\label{sec:2}

In this study we considered stars in the field of view (FOV) of the \textit{Kepler} telescope for which \cite{2014ApJS..211...24M} could determine rotation periods using the auto-correlation function (hereinafter, the ACF method). To ensure that the main source of the variability for the selected stars is magnetic activity, we only selected stars on the main-sequence, using $T_{\rm eff}$ and $\log\,{\rm g}$ values from the \cite{Huber2014} catalogue to exclude giants (see Table\,\ref{tab:1}). We note that \cite{Huber2014} calibrated effective temperatures to the infrared flux temperature scale. This resulted in approximately a 200~K offset from the original Kepler Input catalogue (KIC) \cite[][]{2012ApJS..199...30P}. We also precluded stars flagged in the KIC as giants~(GS), eclipsing binary~(EB), or host stars with planetary transits confirmed~(PTC), with planet candidates~(PC), and false-positive planets~(FP). 

\begin{figure}
\includegraphics[width=\linewidth]{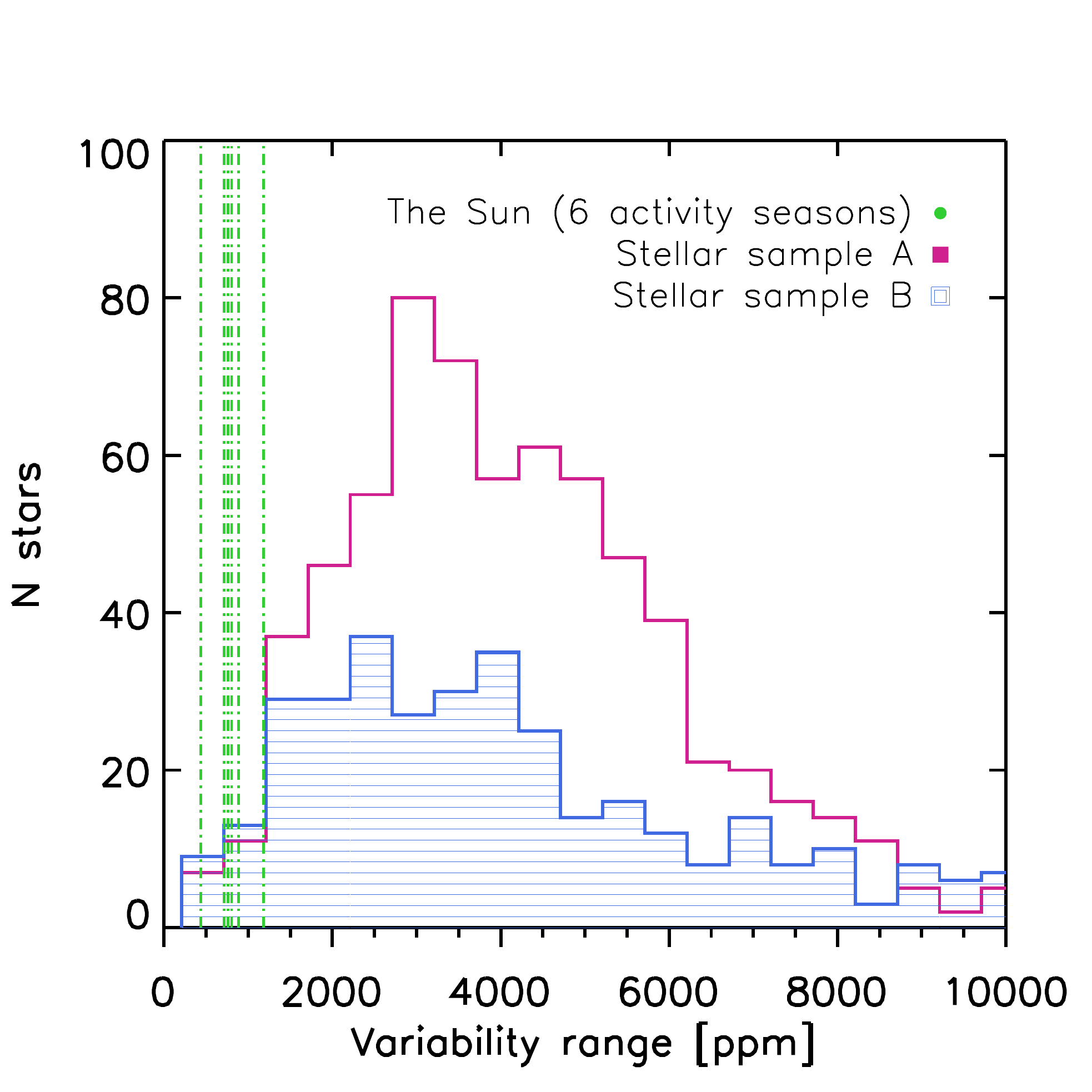}
%\centering
\caption{Histograms for variability ranges from samples A (violet outline) and B (blue rectangles). Vertical green dashed lines represent solar variability calculated for 6 activity seasons using 21 years of VIRGO TSI data from Paper\,II (see text for more details). 
}
\label{fig2}
\end{figure}

We selected two sets of stars with near solar parameters. The selection criteria for both samples (A and B) are illustrated in Fig.\,\ref{Fig1_GPS_III} and given in Table\,\ref{tab:1}. Figure\,\ref{Fig1_GPS_III} and \ref{fig2} show the variability ranges of the set of selected stars. Kepler observatory provided 4~years of photometric information, from 2009 to 2013, segmented in 18 quarters ($Q_{0}-Q_{17}$) due to the telescope reoriented itself every 90~days. The \textit{Kepler} observing quarters resulted in $Q_{0}$ 10 days, and $Q_{1}$ 33 days for the commissioning phase and, segmented 90~days light curves for $Q_{2}$ to $Q_{16}$ \cite[see public data release 25,][]{2016ksci.rept....3T,2016ksci.rept....1V}. The second month of $Q_{17}$ was terminated after less than 5 days observing, after the reaction wheel 4 failed.

Sample\,A (see Figure\,\ref{Fig1_GPS_III} panel\,II) was selected to test the performance of the GPS method for stars with near-solar rotation periods. Hence, in this sample, we considered stars with a narrow range of rotational periods between 24.0 and 27.4~days (i.e. encompassing the sidereal Carrington rotation period of the Sun at 25.4 days) and a broad range of effective temperatures $T_{\rm eff}$\,$\in$\,(5000--6000)~K. These selection criteria yielded a sample consisting of 686~stars. From this sample, 282~stars also have rotation periods obtained by \cite{2015A&A...583A..65R} using the Generalised Lomb-Scargle periodograms (hereinafter, GLS method).

Sample\,B (see Figure\,\ref{Fig1_GPS_III}\,panel\,III) was selected to study the dependence of the inflection point position on the rotation period. Therefore, in contrast to Sample\,A, we considered stars with a broad range of rotation periods (between 15 to 40~days) but a narrow range of effective temperatures (5700--5900\,K, encompassing the solar value of 5778\,K). These criteria led to the selection of 361~stars for Sample\,B after removing overlapping targets the with initial Sample\,A. Rotation periods of 172 stars in this sample were also reported in \cite{2015A&A...583A..65R}.

Between both samples, we thus considered 1047 \textit{Kepler} stars in all. The light curves have been acquired in the long-cadence mode (i.e. with a cadence of 29.42~min). Following \cite{2014ApJS..211...24M} and \cite{2015A&A...583A..65R}, we utilised light curves from $Q_{1}-Q_{14}$ processed with the pre-search conditioning and Bayesian maximum a posteriori approach \cite[PDC-MAP, see][]{2012PASP..124.1000S}. For quarters $Q_{15}-Q_{17}$ only processing with  multiscale MAP \cite[PDC-msMAP][]{2014PASP..126..100S} is available.

In Figure\,\ref{fig2} we plot the distribution of variability ranges in our samples\,A and B. These variability values are defined by computing the difference between the 95th and 5th percentiles of the sorted flux values for each of the \textit{Kepler} observing quarters \citep[see][]{Basri2011} and then taking the median value among the quarters. This defined variability range was chosen versus the approaches based on the standard deviation analysis by \cite{2014JSWSC...4A..15M,2015ApJS..221...18H} or, the smoothed amplitude (10th to 90th) method presented in \cite{2017ApJ...842...83D}, given the higher range of amplitude used in \cite{Basri2011}. The selection of the methods mentioned are not expected to compromise the analysed outcome. Additionally, we show solar variability ranges computed using total solar irradiance data (TSI, i.e. total radiative flux from the Sun at 1\,A.U.) for 1996--2017 obtained by the \textit{Variability of solar IRradiance and Gravity Oscillations} \citep[VIRGO;][]{1997SoPh..175..267F} experiment on the \textit{SOlar and Heliospheric Observatory} SoHO mission. For these VIRGO data, the entire 1996--2017 observation period was split into 6~Kepler-like time ranges (five 1530-day periods and one 712-day period for a total of 7787-days starting 28 January 1996, see Fig.~3 from Paper\,II). The solar variability value for each of the time ranges is represented in Fig.\,\ref{fig2} by vertical green dashed lines. This gives a range of solar variability of $Var_{\odot}\in$\,(400--1300)\,ppm.

\begin{figure*}[!ht]
\includegraphics[width=1.0\textwidth]{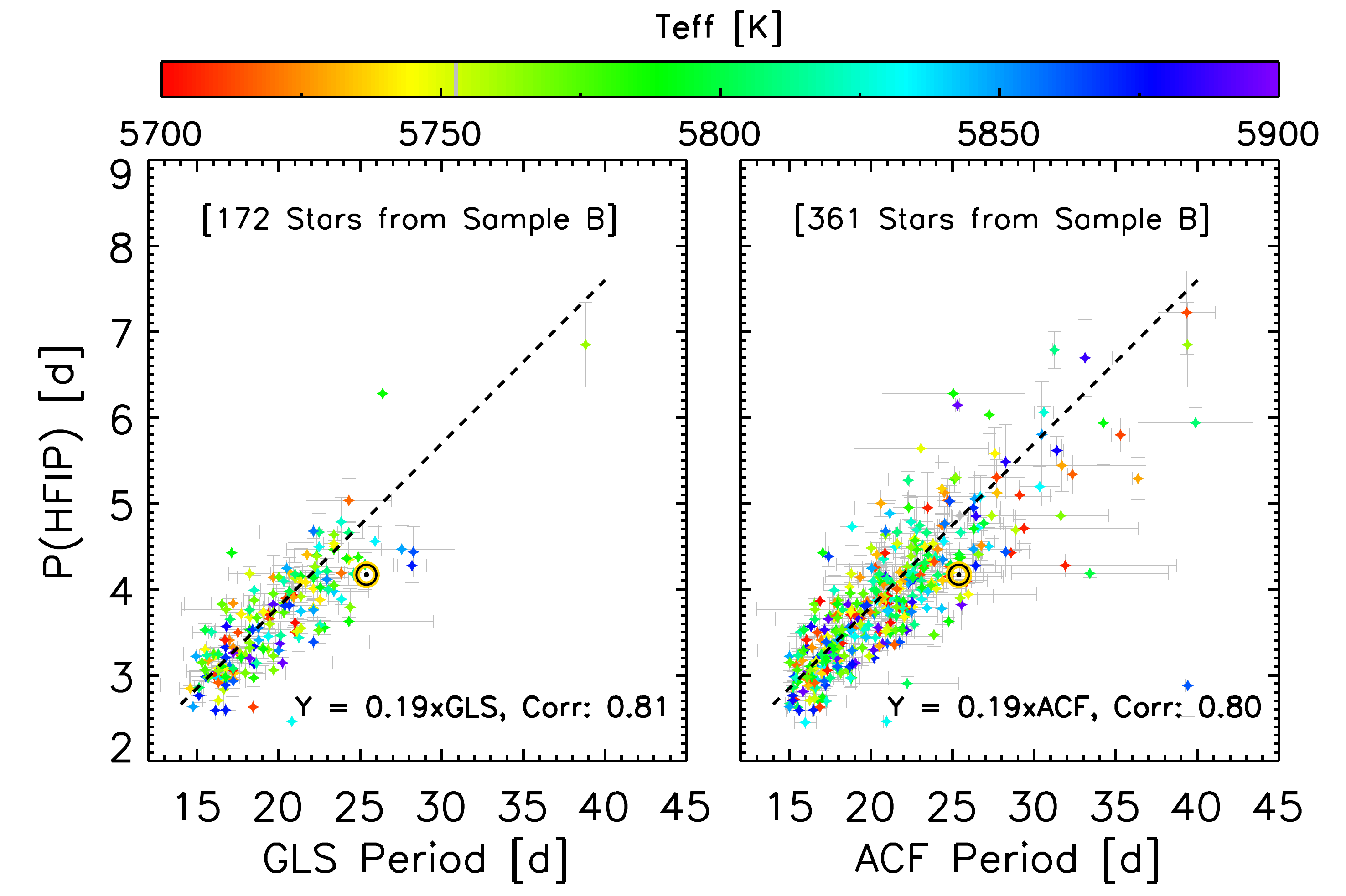}
%\centering
\caption{Only sample B is shown here. Position of the high-frequency inflection point ($P_{\rm HFIP}$) is plotted against rotation period. Rotation periods are taken from \cite{2015A&A...583A..65R} (left panel) and \cite{2014ApJS..211...24M} (right panel). Colours represent the stellar effective temperature, $T_{\rm eff}$. The Sun is represented by the solar symbol~$\odot$. Dashed lines in both panels indicate a linear fit constrained to go through the origin of the coordinate system. A logarithmic visualization is available in the appendix, see Fig.~\ref{fig3_Log}.}\label{fig3}% 3 sources 
\end{figure*}
%@arxiver{Fig3.pdf}

Figure\,\ref{fig2} shows that most of the stars in our samples are much more variable than the Sun. This agrees with  \cite{Garcia2014,2016JSWSC...6A..38B,Reinhold518}, which showed that solar-type stars (i.e. stars with near-solar fundamental parameters and rotation periods) are on average significantly more variable than the Sun. Furthermore, our samples A and B also contain stars which are cooler and rotate faster than the Sun. These stars are also expected to be more variable than the Sun \citep[see, e.g.][for the dependence of the variability on the rotation period and temperature]{2014ApJS..211...24M}.

We note that anomalously low variability of solar-type stars found by \cite{Reinhold518} does not necessarily imply that the Sun is an outlier. An alternative explanation is that by comparing solar variability to the sample of stars with {\it known} rotation periods, we focus only on a small sub-sample of stars for which the ACF method could return rotation periods (and the Sun most probably would not belong to such a sample). Along this line, \cite{Reinhold518} found that solar levels of photometric variability are typical for stars having near-solar fundamental parameters but unknown rotation periods. 

\section{Results and discussion}\label{sec:3}

\begin{table*}
\caption{GPS outcome values.}
\begin{tabular}{l c c c c c c c c c c r}
\hline 
%& & & & & & & & & & & \\[-5pt]
\hline 
& & & & & & & & & & & \\[-2pt]
    \multicolumn{5}{c}{$[  ----------  (1)  ----------   ]$} & \multicolumn{1}{c}{$[ -- (2) -- ]$} & \multicolumn{2}{c}{$[  ---  (3) ---   ]$} & \multicolumn{4}{c}{$[  -----  (4) -----   ]$} \\
& & & & & & & & & & & \\[-2pt]
 KIC &	$P_{\rm HFIP}$ & $\sigma\,P_{\rm HFIP}$ & $\alpha$ & $\sigma\,\alpha$ & $P_{\rm rot}$ GPS & $P_{\rm rot}$ GLS & $P_{\rm rot}$ ACF & Var & $\log\,{\rm g}$ & [Fe/H] &  $T_{\rm eff}$ \\ 
 & [d] & [d] & & & [d] & [d] & [d] & [ppm] &  &  & [K] \\
 & & & & & & & & & & & \\[-2pt]
 \hline
 \hline 
& & & & & & & & & & & \\[0.1pt]
 10070928 & 3.78 & 0.108 & 0.173 & 0.0050 & 19.894 &  22.132 & 21.747 & 4.594 & -0.46 & 4688 & 5706 \\
 10080186 & 3.67 & 0.131 & 0.207 & 0.0074 & 19.315 &  18.239 & 17.747 & 4.547 & -0.10 & 7472 & 5749 \\
 10080239 & 2.96 & 0.084 & 0.184 & 0.0052 & 15.578 &  16.577 & 16.131 & 4.547 & -0.14 & 7174 & 5792 \\
 10083970 & 3.08 & 0.097 & 0.188 & 0.0059 & 16.210 &  16.308 & 16.374 & 4.559 & -0.20 & 11418 & 5745 \\
 10089777 & 3.65 & 0.066 & 0.188 & 0.0034 & 19.210 &  19.437 & 19.381 & 4.541 &  0.07 & 3752 & 5713 \\
 10091612 & 2.90 & 0.078 & 0.130 & 0.0035 & 15.263 &  --	 & 22.214 & 4.550 & -0.18 & 1760 & 5804 \\
 10125510 & 3.78 & 0.180 & 0.161 & 0.0076 & 19.894 &  --	 & 23.427 & 4.372 & -0.64 & 0608 & 5838 \\
 10129857 & 4.69 & 0.235 & 0.162 & 0.0081 & 24.684 &  --	 & 28.859 & 4.536 & -0.06 & 1488 & 5757 \\
 10136417 & 4.46 & 0.276 & 0.169 & 0.0105 & 23.473 &  27.541 & 26.288 & 4.303 &  0.16 & 2690 & 5849 \\
 10140949 & 4.12 & 0.127 & 0.182 & 0.0056 & 21.684 &  --	 & 22.636 & 4.501 &  0.02 & 1999 & 5874 \\
 10146308 & 3.69 & 0.191 & 0.174 & 0.0090 & 19.421 &  --	 & 21.193 & 4.591 & -0.52 & 3340 & 5804 \\
 10064358 & 3.82 & 0.078 & 0.231 & 0.0047 & 20.105 &  16.526 & 16.533 & 4.486 & -0.22 & 2526 & 5772 \\
& & & & & & & & & & & \\[0.1pt]
\hline 
\end{tabular} 
\vspace{0.4cm} 
%\captionsetup{labelformat=empty}
\centering
\tablefoot{This table contains an example of the GPS outputs, the compared rotation period values from GLS \& ACF, and stellar parameters for 12 randomly selected objects from samples A \& B. 1) GPS outcome: In column~2 $P_{\rm HFIP}$ is given, in column~3 its 2-sigma uncertainty, $\sigma\,P_{\rm HFIP}$, defined from individual inflection points for each \textit{Kepler} observing quarter. In column~4 and 5 values of $\alpha$-factor and its 2-sigma uncertainty are reported respectively. $P_{\rm rot}$~GPS values in column~6, as result of applying Eq.~\ref{eq1} using the factor $\alpha=0.19$. 2) Column~7 shows the $P_{\rm rot}$ reported by~\cite{2015A&A...583A..65R}. 3) $P_{\rm rot}$ and variability values reported by~\cite{2014ApJS..211...24M} in column~8. 4) Columns~8, 9 and 10 show the $T_{\rm eff}$, $\log\,{\rm g}$, and [Fe/H] respectively, taken from \cite{Huber2014}. A complete table for the 1047 objects is available in a machine-readable form in the online journal and at the Centre de Données astronomiques de Strasbourg (CDS) -- VizieR Online Data catalogue.}\label{tab:2} 
\end{table*}   

\begin{figure*}[!ht]
\includegraphics[width=1.0\textwidth]{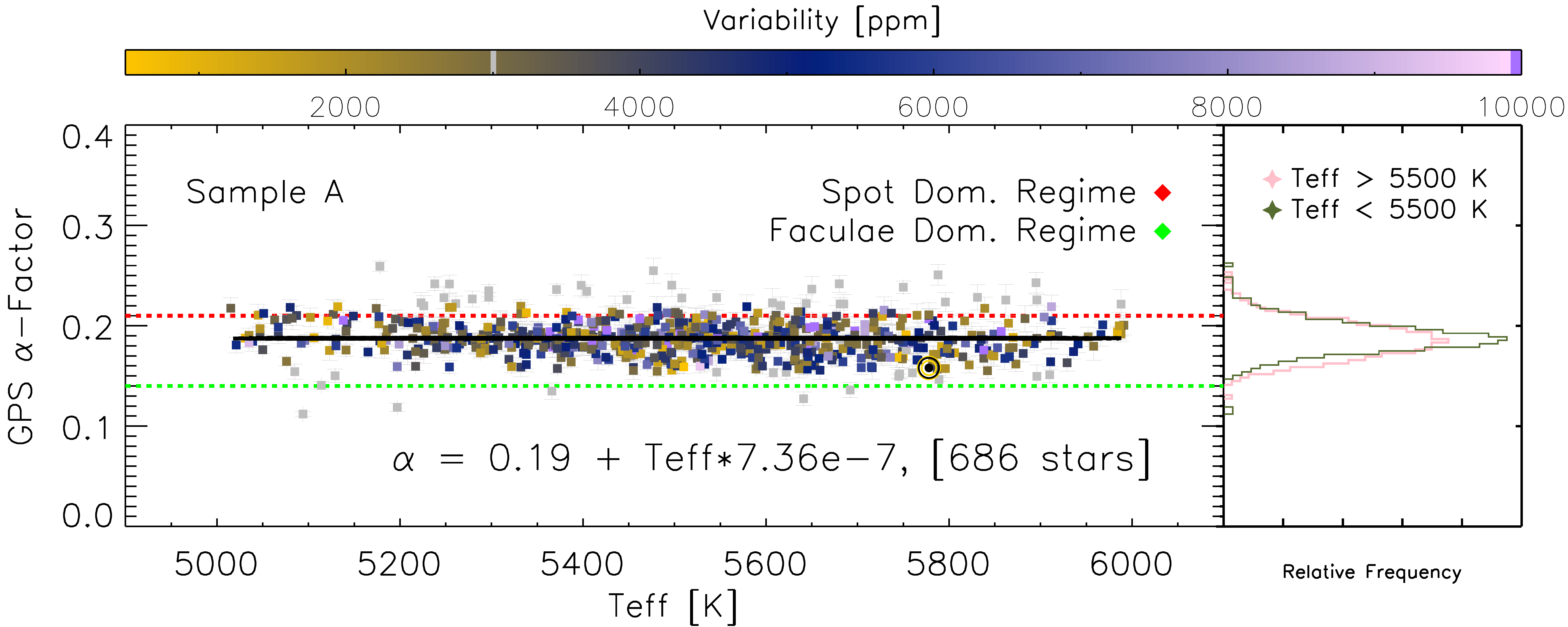}
\includegraphics[trim={0 0 0 40 cm},clip,width=1.\textwidth]{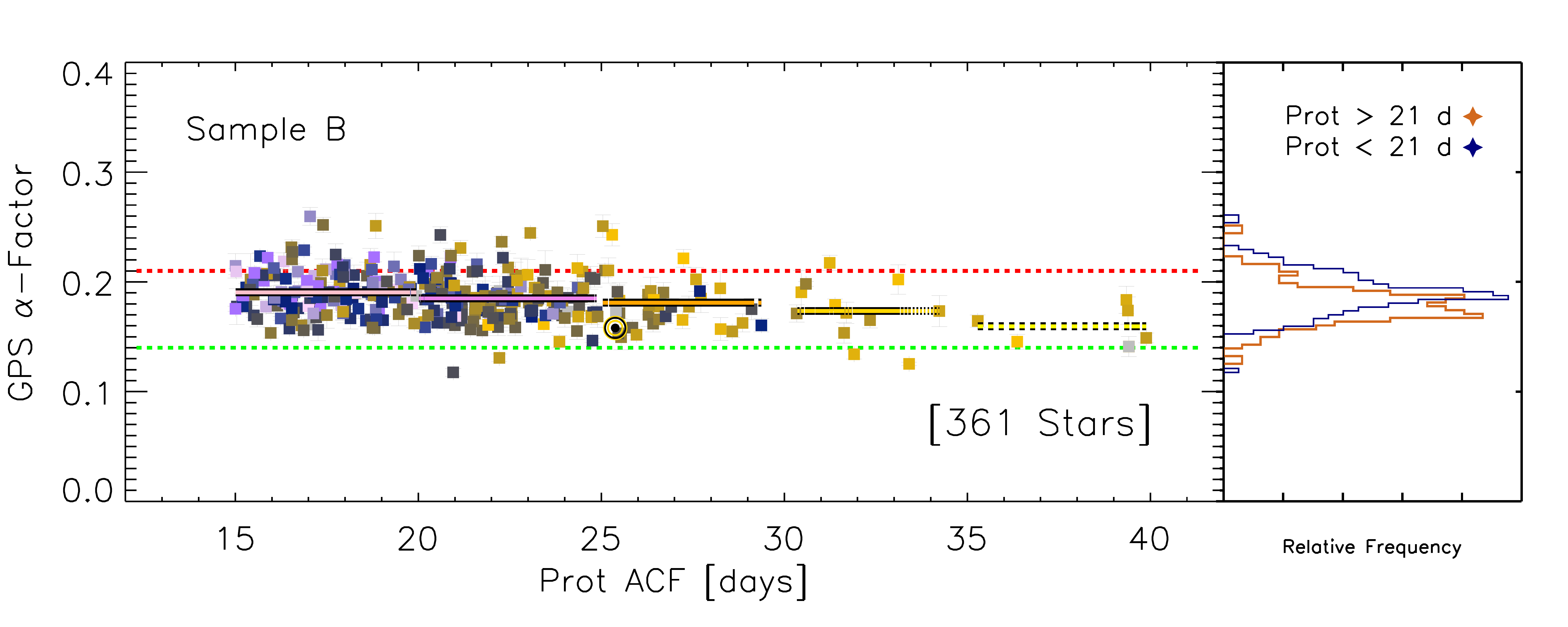}
%\vspace{-0.8cm}
\caption[]{Top panel: $\alpha$\,factor versus $T_{\rm eff}$ for Sample\,A shows consistency across a broad temperature range. The black line corresponds to the linear fit to values with an uncertainty within 2-$\sigma$ of the mean of the distribution as shown by coloured squares (grey squares lie outside the 2-$\sigma$ of the distribution). The histograms to the right side of the panel display the distribution of $\alpha$ values for two effective temperature regimes, using $T_{\rm eff}=5500$\,K as a threshold. Bottom panel: $\alpha$\,factor versus $P_{\rm rot}$ from \cite{2014ApJS..211...24M} for Sample\,B shows a slight decrease in $\alpha$\ with rotation period. The coloured segments indicate the mean of $\alpha$\ for the different $P_{\rm rot}$ ranges as indicated in Table~\ref{tab:3}.
The histograms to the right of the panel indicate the distribution of $\alpha$ values for two rotation period regimens, using 21\,d as a threshold. For both panels the error bars represent 2-$\sigma$ uncertainties of the $\alpha$ values over all \textit{Kepler} quarters available for each star. The gray squares lie outside of a 2-$\sigma$ of the distribution. The dashed red and green horizontal lines represent the $\alpha$\,factor values in the extreme cases with all variability being due to spots ($\alpha=0.21$) and all due to faculae~($\alpha=0.14$), respectively~\footnotemark.}
\label{fig4_5}
\end{figure*}

\noindent In this section, we calculate the position of the inflection point for each star in the samples\,A and\,B defined in Sect.\ref{sec:2}. Following the methodology described in Papers\,I and\,II, we first calculate the power spectra of the stellar brightness variations using a Paul wavelet of order six \citep[see][]{1998BAMS...79...61T} for the \textit{Kepler} observing quarters $Q_{1}-Q_{17}$. We determined the period corresponding to the high-frequency inflection point, $P_{\rm HFIP_{(Q_{n})}}$ per quarter, and calculate the mean value $P_{\rm HFIP}$ over all 17 quarters for each star. This allows us to obtain a unique representative value of $P_{\rm HFIP}$ per star. The uncertainty is calculated using 2-$\sigma$ of the distribution of the obtained $P_{\rm HFIP}$ values. Finally, we used the $P_{\rm HFIP}$ to calculate the stellar rotation period $P_{\rm rot}$ (see Table\,\ref{tab:2} and on-line reference for a compilation of GPS outputs and comparison with GLS and ACF reference values).

In Figure\,\ref{fig3} we plot the mean values of the $P_{\rm HFIP}$ positions for each of the stars against the rotation periods from \cite{2015A&A...583A..65R} (left panel, GLS) and \cite{2014ApJS..211...24M} (right panel, ACF). The rotation periods and positions of the inflection points are well correlated. A linear fit constrained to go through the origin of the coordinate system gives $P_{\rm HFIP} = 0.19\,\times\,P_{\rm rot}$ with Pearson coefficients of 0.81 and 0.80 for periods from \cite{2015A&A...583A..65R} and from \cite{2014ApJS..211...24M}, respectively.

The scatter around the linear fits has multiple sources. First, the calibration coefficient between rotation period and inflection point, $\alpha = P_{\rm HFIP} / P_{\rm rot}$, depends on the relative roles that bright faculae and dark spots play in generating stellar brightness variations. According to the model presented in Paper\,I, these roles are regulated by the ratio between facular and spot areas of active regions at the time of emergence, $S_{\rm fac} / S_{\rm spot}$ (i.e. zero ratio would lead to a purely spot-dominated star, while very large ratios would correspond to a faculae-dominated star). Second, there is an intrinsic statistical uncertainty of the GPS method. For example, in Paper\,I we found that even for a star with a fixed $S_{\rm fac} / S_{\rm spot}$ ratio the factor $\alpha$ showed 5-10\% variations from one realization of active regions emergence to another. Finally, there is also an uncertainly in the determination of rotation periods by \cite{2015A&A...583A..65R} and \cite{2014ApJS..211...24M} (see, e.g., Fig.\,\ref{fig1_App}, where we compare the periods from these two sources for the 172 stars of sample\,B that are common to both).

In Fig.~\ref{fig4_5} we show calibration factors, $\alpha$, for samples\,A (top panel) and\,B (bottom panel). The rotation periods of stars in both samples have been taken from \cite{2014ApJS..211...24M}. In Paper\,I, we demonstrated that the profile of the high-frequency tail of the power spectrum and, consequently, the values of $\alpha$ depend on the center-to-limb variations (CLVs) of the brightness contrasts of magnetic features. Since spots and faculae have different CLVs, the value of $\alpha$ depends on their relative contributions to the stellar brightness variations. For the extreme cases, we found that $\alpha$ is about 0.14 for simulated stellar light curves with variability solely determined by faculae and about 0.21 for simulated stars with variability dominated by spots. These values are respectively designated by the red and green horizontal dashed lines in Fig.\,\ref{fig4_5}. It is reassuring to see that most of the $\alpha$ values for samples A and B appear between these two extreme-cases. Stars with values of $\alpha$ outside of this range (in particular, with $\alpha>0.21$) are likely due to: inclination angles below 45$^{\circ}$, which can lead to a shift of the inflection point to lower frequencies (see Fig.~9 from Paper\,I); statistical noise of the GPS method; and possible uncertainties in rotation periods from \cite{2014ApJS..211...24M}.

For Sample\,A, the ratios are shown as a function of stellar effective temperature from \cite{Huber2014}, while for Sample\,B, they are plotted as a function of stellar rotation period from \cite{2014ApJS..211...24M}. The upper panel of Fig.\,\ref{fig4_5} shows that for near-solar rotation periods (the rotation periods in Sample\,A were constrained between 24 and 27.4 days, see Table\,\ref{tab:1}), the position of the inflection point shows no significant dependence on the effective temperature (e.g. the fitting of a slope gives a value of $7.36\times10^{-7}$, which is well below the $1~\sigma$ uncertainty of $1.8\times10^{-6}$). We also note that the mean value of $\alpha=$~0.19 is equal to the slope of the regression shown in Fig.~\ref{fig3}. This implies that neither the $S_{\rm fac} / S_{\rm spot}$ value nor CLVs of facular and spot contrast change significantly within the 5000-6000~K domain of Sample A. We note, however, that we cannot conclusively exclude the improbable scenario that the effect from the change of the facular and spot contributions to brightness variability on $\alpha$ is compensated by a change of facular and spot CLVs such that the net effect on the inflection point is very small.

The bottom panel of Fig.\,\ref{fig4_5} shows that for stars with near-solar effective temperatures there is a rather weak but statistically significant dependence of the $\alpha$\,factor on the rotation period. For example, fitting of a linear dependence returns a slope value of $9.3\times10^{-4}$ which is 3.8 times larger than its $1\sigma$ uncertainty of $2.5\times10^{-4}$. However, the value of the slope is strongly affected by a couple of slowly rotating stars and, thus, might not represent a trend in the full sample. To better characterise such a trend we calculated the mean value of the calibration factor in several bins of rotation period values. We compiled the mean $\alpha$ values per several bins of rotation periods, see Table~\ref{tab:3}. To further illustrate the trend of $\alpha$ values with rotation period, the histogram to the right side of the panel shows the distributions of $\alpha$ values for two rotation periods - one for stars with rotation periods below 21 days and another with rotation periods above 21 days. One can see that the two distributions are clearly shifted relative to each other and the $\alpha$-values of faster rotating stars are larger than those of the slower-rotating stars.

\begin{table*}
\centering
\caption{Mean $\alpha$-values in sample B per bin.}\label{tab:3}
\begin{tabular}{l c c c c c l}
\hline
\hline
    &           &                      &                &  &\\[-4pt]
Bin & n & $P_{HFIP}$~[d] & $\bar{\alpha}$ & $\sigma$ & $\sigma/\sqrt{n}$ & Bin colour \\ 
  &      &        &        &       &            & \\[-3pt]
\hline
\hline
  &      &           &        &       &             &\\[-1pt]
1 & 158  & [14--20] & 0.190  & 0.001 &$9.0\times10^{-5}$ & \textcolor{CarnationPink}{Pink}\\
2 & 148  & [20--25] & 0.184  & 0.001 &$9.8\times10^{-5}$ & \textcolor{Purple}{Purple}\\
3 & 36   & [25--30] & 0.181  & 0.003 &$5.0\times10^{-4}$ & \textcolor{Orange}{Orange}\\
4 & 12   & [30--35] & 0.173  & 0.006 &$1.9\times10^{-3}$ & \textcolor{Dandelion}{Gold}\\
5 & 6    & [35--40] & 0.156  & 0.004 &$0.17\times10^{-3}$& \textcolor{Goldenrod}{Yellow}\\
  &      &           &        &       &             &\\[-1pt]
\hline
\end{tabular}
\vspace{0.4cm} 
%\captionsetup{labelformat=empty}
 \tablefoot{ \hspace{3cm} Compilation of mean $\alpha$-values for n stars per range of rotation periods, see Fig.\,\ref{fig4_5}}.
\end{table*}

\footnotetext{A logarithmic visualization of the relation $\alpha$\,factor versus $P_{\rm rot}$ is available in the appendix, see Fig.~\ref{fig4_5_Log}.}

\begin{figure*}[!ht]
\includegraphics[width=1.0\textwidth]{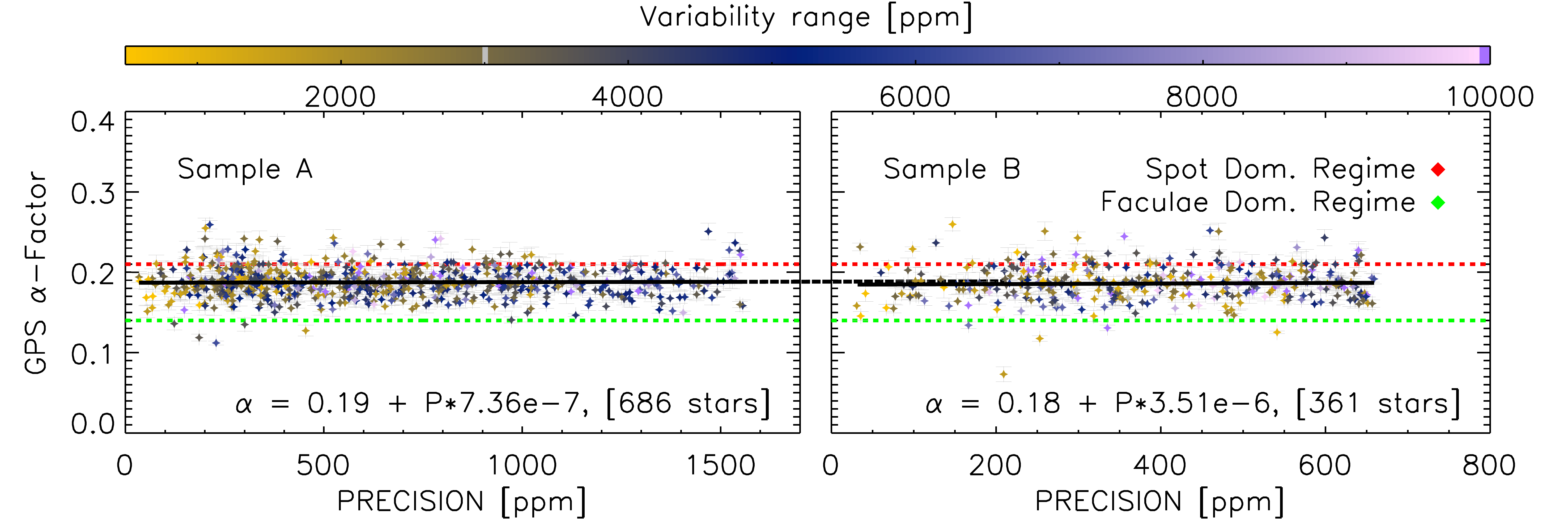}
\caption{$\alpha$\,factor versus photometric precision for sample~A (left panel) and sample~B (right panel) for \textit{Kepler} observations. Those records for single long observation of isolated stars observed in an uncrowded pixel, i.e light curves of resolved targets without contamination of additional sources. %F-, G-, and K-type stars.
Extreme-case limits for spot- and faculae-dominated stars are shown as horizontal dashed lines in red and green, respectively. The individual data points are coloured according to the detected variability range for that particular star. This is consistent with previous figure~\ref{fig4_5}. The error bars represent 2-$\sigma$ uncertainties of the $\alpha$ values distribution over all \textit{Kepler} quarters available per star. Gray rhomboids represent data points that lie more than 2-$\sigma$ from the centre of the distribution.}
\label{fig6}
\end{figure*}

We note that the n number of stars and the amplitude of photometric variability in our samples decreases with rotation period. Consequently, slow rotators might be more affected by photometric noise. We investigated the possible effect of the {\it Kepler} white noise on the deduced positions of inflection points for the stars in our samples. In Fig.\,\ref{fig6}, we plot the dependence of the $\alpha$\,factor values on the expected {\it Kepler} noise levels for each of the stars, calculating the amplitude of the {\it Kepler} noise as a function of the {\it Kepler} magnitude \citep[following][]{2013oepa.book.....L}. The derived \textit{precision} (called the noise in the {\it Kepler} context) accounts for noise introduced by the instrument and gives it as a function of {\it Kepler} magnitude of the source and the variability of sources (see Fig~\ref{RVAR_KEPMAG}). The 99.9\% of the stars in our sample present a {\it Kepler} magnitude of 16~mag or fainter. We find that values of the $\alpha$\,factor are independent of the {\it Kepler} noise, with fits of a linear dependence to samples\,A and B giving slope values well below their $1\sigma$ uncertainties ($7.4\times10^{-7}$ and $4.9\times10^{-8}$, respectively). Consequently, we do not expect the {\it Kepler} noise to affect the positions of inflection points determined for stars in our samples. Furthermore, we note that photometric noise would shift the position of the inflection point to lower frequencies (see Paper\,II), i.e. it would lead to a trend opposite to what is seen in the bottom panel of Fig.\,\ref{fig4_5}. 

A possible explanation of the observed tendency in Fig.\,\ref{fig4_5} is a change of relative contribution of faculae and spots to stellar rotation variability (or $S_{\rm fac} / S_{\rm spot}$ ratio in terms of Paper\,I) with rotation period. The increase of the $\alpha$\,factor with rotation rate implies that the $S_{\rm fac} / S_{\rm spot}$ ratio (and, consequently, the contribution of faculae to the rotational variability) is lower in faster rotating and, therefore, more active stars. Such a trend is consistent with an extrapolation to higher activities of observed solar behaviour. Indeed,  the mean size of spots on the Sun increases during periods of high solar activity \citep{Hathaway_LR,Mandal2020}. At the same time the $S_{\rm fac} / S_{\rm spot}$ ratio decreases with the size of active regions and their spot components. 

An extrapolation of these trends to activity levels higher than seen in the Sun results in an increase of the $\alpha$\,factor with activity, and, consequently, with rotation rate, as indicated by the bottom panel of Fig.\,\ref{fig4_5}. 

We note that the ratio $S_{\rm fac}/S_{\rm spot}$ between facular and spot areas of the individual magnetic features at the moment of their emergence discussed until now is different from the ratio between {\it instantaneous} stellar disk coverage by faculae and spots. The former is a property of a magnetic feature during its emergence onto the surface of the star, while the latter is strongly affected by the evolution of the magnetic flux after emergence. For example, in the hypothetical case of facular portions of active regions evolving exactly as spot portions, these two ratio remain the same. In reality, the {\it instantaneous} ratio is generally significantly larger than that {\it at the time of emergence} since faculae live longer than spots.

Solar observations show that the ratio between {\it instantaneous} solar-disk coverage by faculae and spots decreases as solar activity increases \citep{Chapman1997,foukal1998}.
The observed patterns of stellar-brightness variability indicate that this trend also extends to activity values significantly higher than those observed on the Sun \citep{2014A&A...569A..38S}. Our result indicates that not only the ratio between {\it instantaneous} facular and spot disc coverage shows this trend. Also facular to spot area ratio corresponding to individual active regions {\it at the time of emergence} continues to decrease with increasing level of activity, also  beyond the level of solar activity observed until now. We note that this result is not a simple consequence of the drop of the {\it instantaneous} ratio. Simulations with a surface flux transport model by 
\cite{Cameron_SFTM} show that the origin of the decrease in the {\it instantaneous} ratio with increasing activity is pretty complex. It is, to a large extent, caused by a stronger cancellation of small-scale magnetic field associated with faculae. Consequently, it does not necessarily demand any changes in the structure of the emerging magnetic flux which defines the ratio corresponding to the individual active region {\it at the time of emergence}, see discussion in paper I.

The bottom panel of Fig.\,\ref{fig4_5} shows that the dependence of $\alpha$ on the rotation period is quite noisy, i.e. there is quite a large spread of values for a fixed rotation period. This spread basically covers the entire range of values between faculae- and spot-dominated variability. In particular, it is significantly larger than statistical noise in the inflection point position found in Paper\,I. We speculate that such a large spread implies that the $S_{\rm fac} / S_{\rm spot}$ ratio is not uniquely defined by the stellar effective temperature and rotation period.

In Papers\,I~and\,II, we found that the solar value of the calibration factor ($\alpha_{\rm Sun}=0.158$) is closer to the faculae-dominated case ($\alpha=0.14$) than to the spot-dominated case ($\alpha=0.21$). Interestingly, Fig.\,\ref{fig4_5} shows that the solar ($\alpha_{\rm Sun}$) value appears to be rather low relative to that of stars in both of our samples (see also Fig.\,\ref{fig3}, where the Sun is clearly below the regression line). This is, however, not surprising since most of the stars in our samples are significantly more variable than the Sun even though we selected the stellar sample by the reported detected rotation period and not by the variability \cite[see Fig.\,\ref{fig2}, cf.][]{Reinhold518}.
This implies that these stars are also more active than the Sun \citep[see also][who showed that stars with known near-solar rotation periods have systematically higher values of S-index than the Sun]{Zhang_2020}. Therefore, we can expect that their $S_{\rm fac} / S_{\rm spot}$ ratios are smaller and, consequently, their $\alpha$\,factors are larger. 

\section{Summary}\label{sec:4}

We have developed the GPS method, which is a novel means for determining stellar rotation periods from photometric time series. Instead of basing this determination on the more traditional means of identifying the strongest peak in a Lomb-Scargle periodogram or a maximum of the auto-correlation function, we identify the steepest point (i.e. the inflection point) in the global wavelet power spectrum of stellar brightness variations. In Paper~II, we showed that while the solar brightness contributions from faculae and spots can oppose each other to reduce any peak due to the rotation period from Lomb-Scargle periodograms and auto-correlation functions, it has only a very minor effect on the location of this high-frequency inflection point and the resulting ratio between the period corresponding to the inflection point and the actual rotation period, $\alpha$. In Paper I the factor $\alpha$, however, shows a moderate dependence on the relative contribution of faculae to stellar brightness variations. Therefore, identifying the position of the inflection point allowed determination of the rotation period in stars where other methods fail (with an internal uncertainty of about 25\%). At the same time, this GPS method allows assessing the relative role of faculae in stars with known rotation periods. In Paper~II, we tested the performance of the GPS method against solar photometric data. We demonstrated that, in contrast to other methods, the GPS method allows an accurate determination of the solar rotation period independently of the solar activity level. 

In this study we have applied the GPS method to 1047 F-, G-, and K-type stars with rotation periods reported in \cite{2014ApJS..211...24M}. We have shown that the position of the high-frequency inflection point is well correlated with rotation periods of stars in our two samples analysed, providing further validation of the GPS method. We emphasise that the stellar light curves analysed in this study and the solar light curves analysed in Paper~II are quite different: the amplitudes of brightness variability in the stellar samples in this study are generally higher than that of the Sun, and the stellar brightness modulation are much more regular over rotational timescales.

We find that the $\alpha$ factor increases with rotation rate, indicating that faculae become less important on stars rotating faster than the Sun. We have also found that the facular contribution to solar brightness variability is larger than its contribution to brightness variability in a sample of stars having near-solar rotation periods and temperatures. We attribute this to a selection effect, since the rotation periods of stars with brightness-variability patterns similar to that of the Sun are rather difficult to measure via the ACF method for rotation-period determinations, and thus there is a dearth of such stars in our sample. Consequently, our results indicate that, in addition to being more active than the Sun \citep[see also][]{Reinhold518,Zhang_2020}, the stars with near-solar effective temperatures and near-solar rotation periods determined by \cite{2014ApJS..211...24M} have different compositions of active regions (with smaller facular contributions). The GPS method for determining rotation periods could thus be an important contributor to enhance the lower-than-expected number of G-type stars with near-solar rotation periods reported by \cite{2019ApJ...872..128V}. This method can also improve the solar-stellar comparison as in \cite{Reinhold518}. The outcome of GPS might bring a new perspective in understanding stellar activity.
 
While in this study we focus on applying GPS to stars with known rotation periods in a forthcoming study, we plan to apply the GPS method to {\it Kepler} stars with previously unknown rotation periods as well as to {\it TESS} stars. This might help establish a new and more complete sample of stars having near-solar rotation periods, from which we can investigate whether solar variability still appears anomalously low in comparison to stars in this broader sample.

This will be of importance to the exoplanet community, since knowledge of rotation periods will help identify radial velocity jitter from planetary signals \cite[see][]{2018ASSP...49..239O,2019arXiv191111714F,Hojjatpanah2020}. The anticipated GPS-determined expanded database of stellar rotation periods could also bring crucial information for ongoing and upcoming surveys like NIRPS-HARPS and ESPRESSO \citep[see][]{EXPRESSO,2018EPSC...12.1147B}. Additionally, the precise determination of host-star rotation periods is important for recovering accurate exoplanet radii, which will be crucial for searches of transiting Earths or Super-Earths in future light curves of solar twins in the PLATO field \citep[see][]{PLATO}.

\begin{acknowledgements}

We would like to thank the referee for the constructive comments which helped to improve the quality of this paper. This work was supported by the International Max-Planck Research School~(IMPRS) for Solar System Science at the University of G{\"o}ttingen and European Research Council under the European Union Horizon~2020 research and innovation program (grant agreement by the No. 715947). M.O. acknowledges the support of the Deutsche Forschungsgemeinschft~(DFG) priority program~SPP 1992, Exploring the Diversity of Extrasolar Planets (RE 1664/17-1). E.~M.~A.~G. and M.~O. also acknowledge the support of the FCT/DAAD bilateral grant 2019 (DAAD\,ID:\,57453096). Financial support was also provided by the Brain Korea 21 plus program through the National Research Foundation funded by the Ministry of Education of Korea and by the German Federal Ministry of Education and Research under project 01LG1209A. We would like to thank the International Space Science Institute, Bern, for their support of science team~446 and the resulting helpful discussions. This paper includes data collected by the \textit{Kepler} mission. Funding for the \textit{Kepler} mission is provided by the NASA Science Mission directorate. The data presented were obtained from the Mikulski Archive for Space Telescopes (MAST). STScI is operated by the Association of Universities for Research in Astronomy, Inc., under NASA contract NAS5-26555.
 
\end{acknowledgements}

\bibliographystyle{aasjournal}
\bibliography{Biblio}

\begin{thebibliography}{}
\expandafter\ifx\csname natexlab\endcsname\relax\def\natexlab#1{#1}\fi
\providecommand{\url}[1]{\href{#1}{#1}}

\bibitem[{{Aigrain} {et~al.}(2015){Aigrain}, {Llama}, {Ceillier}, {Chagas},
  {Davenport}, {Garc{\'{\i}}a}, {Hay}, {Lanza}, {McQuillan}, {Mazeh}, {de
  Medeiros}, {Nielsen}, \& {Reinhold}}]{2015MNRAS.450.3211A}
{Aigrain}, S., {Llama}, J., {Ceillier}, T., {et~al.} 2015, \mnras, 450, 3211

\bibitem[{{Amazo-G{\'o}mez} {et~al.}(2020){Amazo-G{\'o}mez}, {Shapiro},
  {Solanki}, {Krivova}, {Kopp}, {Reinhold}, {Oshagh}, \& {Reiners}}]{Eliana1}
{Amazo-G{\'o}mez}, E.~M., {Shapiro}, A.~I., {Solanki}, S.~K., {et~al.} 2020,
  \aap, 636, A69

\bibitem[{{Angus} {et~al.}(2018){Angus}, {Morton}, {Aigrain}, {Foreman-Mackey},
  \& {Rajpaul}}]{2018MNRAS.474.2094A}
{Angus}, R., {Morton}, T., {Aigrain}, S., {Foreman-Mackey}, D., \& {Rajpaul},
  V. 2018, \mnras, 474, 2094

\bibitem[{{Basri}(2018)}]{Basri2018}
{Basri}, G. 2018, \apj, 865, 142

\bibitem[{{Basri} {et~al.}(2011){Basri}, {Walkowicz}, {Batalha}, {Gilliland},
  {Jenkins}, {Borucki}, {Koch}, {Caldwell}, {Dupree}, {Latham}, {Marcy},
  {Meibom}, \& {Brown}}]{Basri2011}
{Basri}, G., {Walkowicz}, L.~M., {Batalha}, N., {et~al.} 2011, \aj, 141, 20

\bibitem[{{Borucki} {et~al.}(2010){Borucki}, {Koch}, {Basri}, {Batalha},
  {Brown}, {Caldwell}, {Caldwell}, {Christensen-Dalsgaard}, {Cochran},
  {DeVore}, {Dunham}, {Dupree}, {Gautier}, {Geary}, {Gilliland}, {Gould},
  {Howell}, {Jenkins}, {Kondo}, {Latham}, {Marcy}, {Meibom}, {Kjeldsen},
  {Lissauer}, {Monet}, {Morrison}, {Sasselov}, {Tarter}, {Boss}, {Brownlee},
  {Owen}, {Buzasi}, {Charbonneau}, {Doyle}, {Fortney}, {Ford}, {Holman},
  {Seager}, {Steffen}, {Welsh}, {Rowe}, {Anderson}, {Buchhave}, {Ciardi},
  {Walkowicz}, {Sherry}, {Horch}, {Isaacson}, {Everett}, {Fischer}, {Torres},
  {Johnson}, {Endl}, {MacQueen}, {Bryson}, {Dotson}, {Haas}, {Kolodziejczak},
  {Van Cleve}, {Chandrasekaran}, {Twicken}, {Quintana}, {Clarke}, {Allen},
  {Li}, {Wu}, {Tenenbaum}, {Verner}, {Bruhweiler}, {Barnes}, \&
  {Prsa}}]{2010Sci...327..977B}
{Borucki}, W.~J., {Koch}, D., {Basri}, G., {et~al.} 2010, Science, 327, 977

\bibitem[{{Bouchy} \& {Doyon}(2018)}]{2018EPSC...12.1147B}
{Bouchy}, F., \& {Doyon}, R. 2018, in European Planetary Science Congress,
  EPSC2018--1147

\bibitem[{{Buzasi} {et~al.}(2016){Buzasi}, {Lezcano}, \&
  {Preston}}]{2016JSWSC...6A..38B}
{Buzasi}, D., {Lezcano}, A., \& {Preston}, H.~L. 2016, Journal of Space Weather
  and Space Climate, 6, A38

\bibitem[{{Cameron} {et~al.}(2010){Cameron}, {Jiang}, {Schmitt}, \&
  {Sch{\"u}ssler}}]{Cameron_SFTM}
{Cameron}, R.~H., {Jiang}, J., {Schmitt}, D., \& {Sch{\"u}ssler}, M. 2010,
  \apj, 719, 264

\bibitem[{{Chapman} {et~al.}(1997){Chapman}, {Cookson}, \&
  {Dobias}}]{Chapman1997}
{Chapman}, G.~A., {Cookson}, A.~M., \& {Dobias}, J.~J. 1997, \apj, 482, 541

\bibitem[{{Douglas} {et~al.}(2017){Douglas}, {Ag{\"u}eros}, {Covey}, \&
  {Kraus}}]{2017ApJ...842...83D}
{Douglas}, S.~T., {Ag{\"u}eros}, M.~A., {Covey}, K.~R., \& {Kraus}, A. 2017,
  \apj, 842, 83

\bibitem[{{Faria} {et~al.}(2019){Faria}, {Adibekyan}, {Amazo-G{\'o}mez},
  {Barros}, {Camacho}, {Demangeon}, {Figueira}, {Mortier}, {Oshagh}, {Pepe},
  {Santos}, {Gomes da Silva}, {Costa Silva}, {Sousa}, {Ulmer-Moll}, \&
  {Viana}}]{2019arXiv191111714F}
{Faria}, J.~P., {Adibekyan}, V., {Amazo-G{\'o}mez}, E.~M., {et~al.} 2019, arXiv
  e-prints, arXiv:1911.11714

\bibitem[{{Foukal}(1998)}]{foukal1998}
{Foukal}, P. 1998, \apj, 500, 958

\bibitem[{{Fr{\"o}hlich} {et~al.}(1997){Fr{\"o}hlich}, {Crommelynck}, {Wehrli},
  {Anklin}, {Dewitte}, {Fichot}, {Finsterle}, {Jim{\'e}nez}, {Chevalier}, \&
  {Roth}}]{1997SoPh..175..267F}
{Fr{\"o}hlich}, C., {Crommelynck}, D.~A., {Wehrli}, C., {et~al.} 1997,
  \solphys, 175, 267

\bibitem[{{Garc{\'{\i}}a} {et~al.}(2014){Garc{\'{\i}}a}, {Ceillier},
  {Salabert}, {Mathur}, {van Saders}, {Pinsonneault}, {Ballot}, {Beck},
  {Bloemen}, {Campante}, {Davies}, {do Nascimento}, {Mathis}, {Metcalfe},
  {Nielsen}, {Su{\'a}rez}, {Chaplin}, {Jim{\'e}nez}, \& {Karoff}}]{Garcia2014}
{Garc{\'{\i}}a}, R.~A., {Ceillier}, T., {Salabert}, D., {et~al.} 2014, \aap,
  572, A34

\bibitem[{{Hathaway}(2015)}]{Hathaway_LR}
{Hathaway}, D.~H. 2015, Living Reviews in Solar Physics, 12, 4

\bibitem[{{He} {et~al.}(2015){He}, {Wang}, \& {Yun}}]{2015ApJS..221...18H}
{He}, H., {Wang}, H., \& {Yun}, D. 2015, \apjs, 221, 18

\bibitem[{{Hojjatpanah} {et~al.}(2020){Hojjatpanah}, {Oshagh}, {Figueira},
  {Santos}, {Amazo-G{\'o}mez}, {Sousa}, {Adibekyan}, {Akinsanmi}, {Demangeon},
  {Faria}, {Gomes da Silva}, \& {Meunier}}]{Hojjatpanah2020}
{Hojjatpanah}, S., {Oshagh}, M., {Figueira}, P., {et~al.} 2020, \aap, 639, A35

\bibitem[{{Huber} {et~al.}(2014){Huber}, {Silva Aguirre}, {Matthews},
  {Pinsonneault}, {Gaidos}, {Garc{\'{\i}}a}, {Hekker}, {Mathur}, {Mosser},
  {Torres}, {Bastien}, {Basu}, {Bedding}, {Chaplin}, {Demory}, {Fleming},
  {Guo}, {Mann}, {Rowe}, {Serenelli}, {Smith}, \& {Stello}}]{Huber2014}
{Huber}, D., {Silva Aguirre}, V., {Matthews}, J.~M., {et~al.} 2014, \apjs, 211,
  2

\bibitem[{{Lammer}(2013)}]{2013oepa.book.....L}
{Lammer}, H. 2013, {Origin and Evolution of Planetary Atmospheres} (Springer
  Berlin Heidelberg), doi:10.1007/978-3-642-32087-3

\bibitem[{{Lanza} \& {Shkolnik}(2014)}]{2014MNRAS.443.1451L}
{Lanza}, A.~F., \& {Shkolnik}, E.~L. 2014, \mnras, 443, 1451

\bibitem[{{Mandal} {et~al.}(2020){Mandal}, {Krivova}, {Solanki}, {Sinha}, \&
  {Banerjee}}]{Mandal2020}
{Mandal}, S., {Krivova}, N.~A., {Solanki}, S.~K., {Sinha}, N., \& {Banerjee},
  D. 2020, arXiv e-prints, arXiv:2004.14618

\bibitem[{{Mathur} {et~al.}(2014){Mathur}, {Salabert}, {Garc{\'\i}a}, \&
  {Ceillier}}]{2014JSWSC...4A..15M}
{Mathur}, S., {Salabert}, D., {Garc{\'\i}a}, R.~A., \& {Ceillier}, T. 2014,
  Journal of Space Weather and Space Climate, 4, A15

\bibitem[{{McQuillan} {et~al.}(2014){McQuillan}, {Mazeh}, \&
  {Aigrain}}]{2014ApJS..211...24M}
{McQuillan}, A., {Mazeh}, T., \& {Aigrain}, S. 2014, \apjs, 211, 24

\bibitem[{{Nielsen} {et~al.}(2013){Nielsen}, {Gizon}, {Schunker}, \&
  {Karoff}}]{2013A&A...557L..10N}
{Nielsen}, M.~B., {Gizon}, L., {Schunker}, H., \& {Karoff}, C. 2013, \aap, 557,
  L10

\bibitem[{{Oshagh}(2018)}]{2018ASSP...49..239O}
{Oshagh}, M. 2018, Asteroseismology and Exoplanets: Listening to the Stars and
  Searching for New Worlds, 49, 239

\bibitem[{{Pepe} {et~al.}(2010){Pepe}, {Cristiani}, {Rebolo Lopez}, {Santos},
  {Amorim}, {Avila}, {Benz}, {Bonifacio}, {Cabral}, {Carvas}, {Cirami},
  {Coelho}, {Comari}, {Coretti}, {De Caprio}, {Dekker}, {Delabre}, {Di
  Marcantonio}, {D'Odorico}, {Fleury}, {Garc{\'{\i}}a}, {Herreros Linares},
  {Hughes}, {Iwert}, {Lima}, {Lizon}, {Lo Curto}, {Lovis}, {Manescau},
  {Martins}, {M{\'e}gevand}, {Moitinho}, {Molaro}, {Monteiro}, {Monteiro},
  {Pasquini}, {Mordasini}, {Queloz}, {Rasilla}, {Rebord{\~a}o}, {Santana
  Tschudi}, {Santin}, {Sosnowska}, {Span{\`o}}, {Tenegi}, {Udry}, {Vanzella},
  {Viel}, {Zapatero Osorio}, \& {Zerbi}}]{EXPRESSO}
{Pepe}, F.~A., {Cristiani}, S., {Rebolo Lopez}, R., {et~al.} 2010, in
  \procspie, Vol. 7735, Ground-based and Airborne Instrumentation for Astronomy
  III, 77350F

\bibitem[{{Pinsonneault} {et~al.}(2012){Pinsonneault}, {An},
  {Molenda-{\.Z}akowicz}, {Chaplin}, {Metcalfe}, \&
  {Bruntt}}]{2012ApJS..199...30P}
{Pinsonneault}, M.~H., {An}, D., {Molenda-{\.Z}akowicz}, J., {et~al.} 2012,
  \apjs, 199, 30

\bibitem[{{Rauer} {et~al.}(2014){Rauer}, {Catala}, {Aerts}, {Appourchaux},
  {Benz}, {Brandeker}, {Christensen-Dalsgaard}, {Deleuil}, {Gizon}, {Goupil},
  {G{\"u}del}, {Janot-Pacheco}, {Mas-Hesse}, {Pagano}, {Piotto}, {Pollacco},
  {Santos}, {Smith}, {Su{\'a}rez}, {Szab{\'o}}, {Udry}, {Adibekyan}, {Alibert},
  {Almenara}, {Amaro-Seoane}, {Eiff}, {Asplund}, {Antonello}, {Barnes},
  {Baudin}, {Belkacem}, {Bergemann}, {Bihain}, {Birch}, {Bonfils}, {Boisse},
  {Bonomo}, {Borsa}, {Brand{\~a}o}, {Brocato}, {Brun}, {Burleigh}, {Burston},
  {Cabrera}, {Cassisi}, {Chaplin}, {Charpinet}, {Chiappini}, {Church},
  {Csizmadia}, {Cunha}, {Damasso}, {Davies}, {Deeg}, {D{\'{\i}}az}, {Dreizler},
  {Dreyer}, {Eggenberger}, {Ehrenreich}, {Eigm{\"u}ller}, {Erikson}, {Farmer},
  {Feltzing}, {de Oliveira Fialho}, {Figueira}, {Forveille}, {Fridlund},
  {Garc{\'{\i}}a}, {Giommi}, {Giuffrida}, {Godolt}, {Gomes da Silva},
  {Granzer}, {Grenfell}, {Grotsch-Noels}, {G{\"u}nther}, {Haswell}, {Hatzes},
  {H{\'e}brard}, {Hekker}, {Helled}, {Heng}, {Jenkins}, {Johansen},
  {Khodachenko}, {Kislyakova}, {Kley}, {Kolb}, {Krivova}, {Kupka}, {Lammer},
  {Lanza}, {Lebreton}, {Magrin}, {Marcos-Arenal}, {Marrese}, {Marques},
  {Martins}, {Mathis}, {Mathur}, {Messina}, {Miglio}, {Montalban}, {Montalto},
  {Monteiro}, {Moradi}, {Moravveji}, {Mordasini}, {Morel}, {Mortier},
  {Nascimbeni}, {Nelson}, {Nielsen}, {Noack}, {Norton}, {Ofir}, {Oshagh},
  {Ouazzani}, {P{\'a}pics}, {Parro}, {Petit}, {Plez}, {Poretti}, {Quirrenbach},
  {Ragazzoni}, {Raimondo}, {Rainer}, {Reese}, {Redmer}, {Reffert},
  {Rojas-Ayala}, {Roxburgh}, {Salmon}, {Santerne}, {Schneider}, {Schou},
  {Schuh}, {Schunker}, {Silva-Valio}, {Silvotti}, {Skillen}, {Snellen}, {Sohl},
  {Sousa}, {Sozzetti}, {Stello}, {Strassmeier}, {{\v S}vanda}, {Szab{\'o}},
  {Tkachenko}, {Valencia}, {Van Grootel}, {Vauclair}, {Ventura}, {Wagner},
  {Walton}, {Weingrill}, {Werner}, {Wheatley}, \& {Zwintz}}]{PLATO}
{Rauer}, H., {Catala}, C., {Aerts}, C., {et~al.} 2014, Experimental Astronomy,
  38, 249

\bibitem[{{Reinhold} \& {Gizon}(2015)}]{2015A&A...583A..65R}
{Reinhold}, T., \& {Gizon}, L. 2015, \aap, 583, A65

\bibitem[{Reinhold {et~al.}(2020)Reinhold, Shapiro, Solanki, Montet, Krivova,
  Cameron, \& Amazo-G{\'o}mez}]{Reinhold518}
Reinhold, T., Shapiro, A.~I., Solanki, S.~K., {et~al.} 2020, Science, 368, 518.
\newblock \url{https://science.sciencemag.org/content/368/6490/518}

\bibitem[{Santos {et~al.}(2019)Santos, Garc{\'{\i}}a, Mathur, Bugnet, van
  Saders, Metcalfe, Simonian, \& Pinsonneault}]{Santos_2019}
Santos, A. R.~G., Garc{\'{\i}}a, R.~A., Mathur, S., {et~al.} 2019, The
  Astrophysical Journal Supplement Series, 244, 21.
\newblock \url{https://doi.org/10.3847\%2F1538-4365\%2Fab3b56}

\bibitem[{{Shapiro} {et~al.}(2020){Shapiro}, {Amazo-G{\'o}mez}, {Krivova}, \&
  {Solanki}}]{paperI}
{Shapiro}, A.~I., {Amazo-G{\'o}mez}, E.~M., {Krivova}, N.~A., \& {Solanki},
  S.~K. 2020, \aap, 633, A32

\bibitem[{{Shapiro} {et~al.}(2017){Shapiro}, {Solanki}, {Krivova}, {Cameron},
  {Yeo}, \& {Schmutz}}]{Shapiro2017}
{Shapiro}, A.~I., {Solanki}, S.~K., {Krivova}, N.~A., {et~al.} 2017, Nature
  Astronomy, 1, 612

\bibitem[{{Shapiro} {et~al.}(2014){Shapiro}, {Solanki}, {Krivova}, {Schmutz},
  {Ball}, {Knaack}, {Rozanov}, \& {Unruh}}]{2014A&A...569A..38S}
---. 2014, \aap, 569, A38

\bibitem[{{Smith} {et~al.}(2012){Smith}, {Stumpe}, {Van Cleve}, {Jenkins},
  {Barclay}, {Fanelli}, {Girouard}, {Kolodziejczak}, {McCauliff}, {Morris}, \&
  {Twicken}}]{2012PASP..124.1000S}
{Smith}, J.~C., {Stumpe}, M.~C., {Van Cleve}, J.~E., {et~al.} 2012, \pasp, 124,
  1000

\bibitem[{{Solanki}(1993)}]{solanki1993}
{Solanki}, S.~K. 1993, \ssr, 63, 1

\bibitem[{{Solanki} {et~al.}(2006){Solanki}, {Inhester}, \&
  {Sch{\"u}ssler}}]{2006RPPh...69..563S}
{Solanki}, S.~K., {Inhester}, B., \& {Sch{\"u}ssler}, M. 2006, Reports on
  Progress in Physics, 69, 563

\bibitem[{{Stumpe} {et~al.}(2014){Stumpe}, {Smith}, {Catanzarite}, {Van Cleve},
  {Jenkins}, {Twicken}, \& {Girouard}}]{2014PASP..126..100S}
{Stumpe}, M.~C., {Smith}, J.~C., {Catanzarite}, J.~H., {et~al.} 2014, \pasp,
  126, 100

\bibitem[{{Thompson} {et~al.}(2016){Thompson}, {Caldwell}, {Jenkins},
  {Barclay}, {Barentsen}, {Bryson}, {Burke}, {Campbell}, {Catanzarite},
  {Christiansen}, {Clarke}, {Colon}, {Coughlin}, {Girouard}, {Haas},
  {Harrison}, {Ibrahim}, {Klaus}, {Li}, {McCauliff}, {Morris}, {Mullally},
  {Rowe}, {Sabale}, {Seader}, {Smith}, {Tenenbaum}, {Twicken}, {Kamal Uddin},
  \& {Van Cleve}}]{2016ksci.rept....3T}
{Thompson}, S.~E., {Caldwell}, D.~A., {Jenkins}, J.~M., {et~al.} 2016, {Kepler
  Data Release 25 Notes}, Kepler Science Document, ,

\bibitem[{{Torrence} \& {Compo}(1998)}]{1998BAMS...79...61T}
{Torrence}, C., \& {Compo}, G.~P. 1998, Bulletin of the American Meteorological
  Society, 79, 61

\bibitem[{{Van Cleve} \& {Caldwell}(2016)}]{2016ksci.rept....1V}
{Van Cleve}, J.~E., \& {Caldwell}, D.~A. 2016, {Kepler Instrument Handbook},
  Tech. rep.

\bibitem[{{van Saders} {et~al.}(2019){van Saders}, {Pinsonneault}, \&
  {Barbieri}}]{2019ApJ...872..128V}
{van Saders}, J.~L., {Pinsonneault}, M.~H., \& {Barbieri}, M. 2019, \apj, 872,
  128

\bibitem[{{Walkowicz} \& {Basri}(2013)}]{Walkowicz2013}
{Walkowicz}, L.~M., \& {Basri}, G.~S. 2013, \mnras, 436, 1883

\bibitem[{{Witzke} {et~al.}(2020){Witzke}, {Reinhold}, {Shapiro}, {Krivova}, \&
  {Solanki}}]{witzke2020}
{Witzke}, V., {Reinhold}, T., {Shapiro}, A.~I., {Krivova}, N.~A., \& {Solanki},
  S.~K. 2020, \aap, 634, L9

\bibitem[{Zhang {et~al.}(2020)Zhang, Shapiro, Bi, Xiang, Reinhold, Sowmya, Li,
  Li, Yu, Du, \& Zhang}]{Zhang_2020}
Zhang, J., Shapiro, A.~I., Bi, S., {et~al.} 2020, The Astrophysical Journal,
  894, L11.
\newblock \url{https://doi.org/10.3847\%2F2041-8213\%2Fab8795}

\end{thebibliography}

\begin{appendix}

%\appendix
\section{A}

\begin{figure}[!ht]
\includegraphics[width=1.0\textwidth]{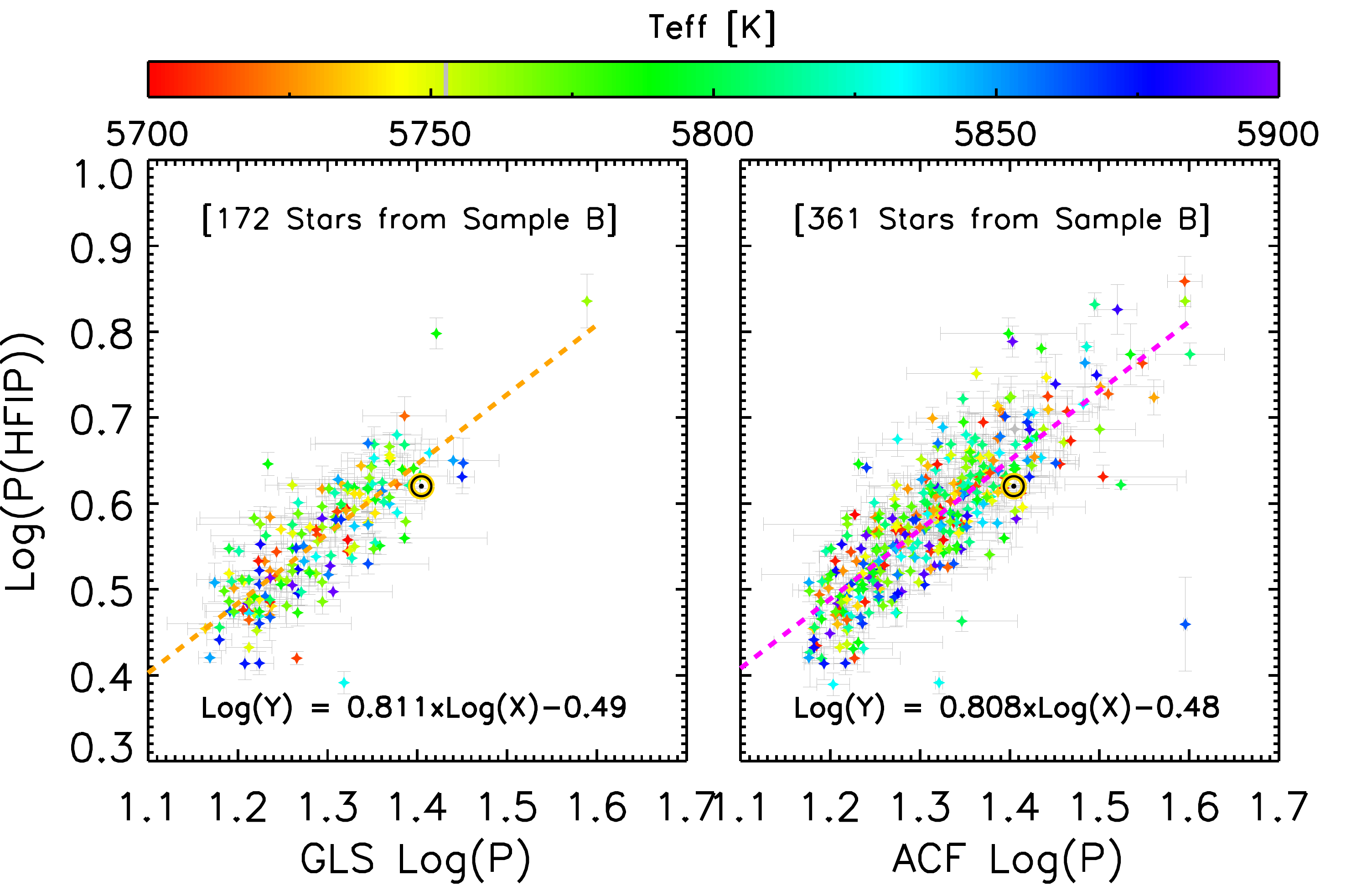}
\caption{Logarithmic visualization of Fig.~\ref{fig3}.}\label{fig3_Log}
\end{figure}

\begin{figure*}[!ht]
\includegraphics[trim={0 0 0 0 cm},clip,width=1.\textwidth]{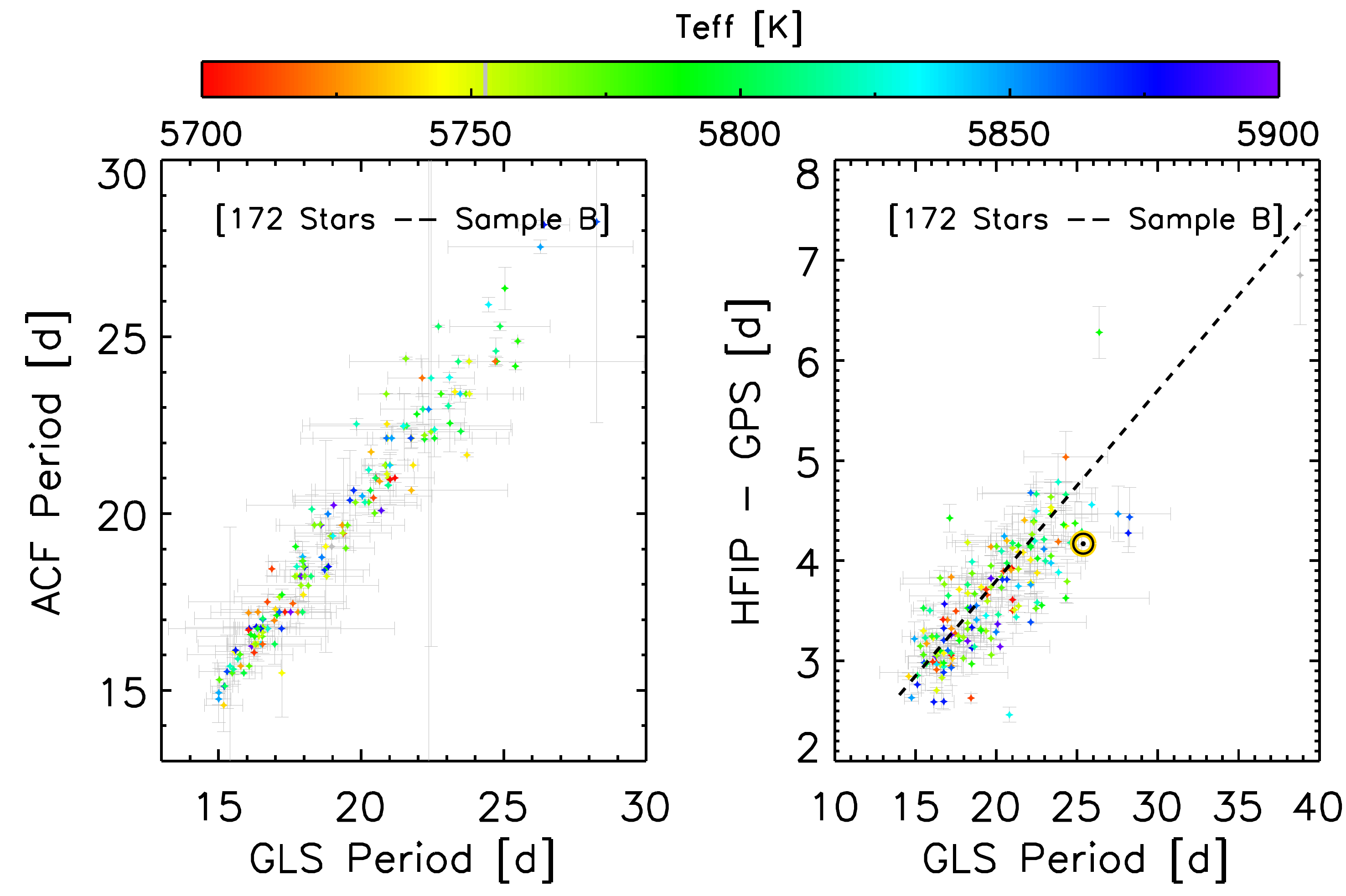}
\caption{Comparison for the 172 stars in sample\,B with reported rotation periods by the ACF, GLS and GPS methods. Left: ACF versus GLS. Right: HFIP-GPS versus GLS. Scatter colour to visualise the temperature. The comparison is made in a similar range scale for a better visualization.}\label{fig1_App}
\end{figure*}

\begin{figure*}[!ht]
\includegraphics[width=1.0\textwidth]{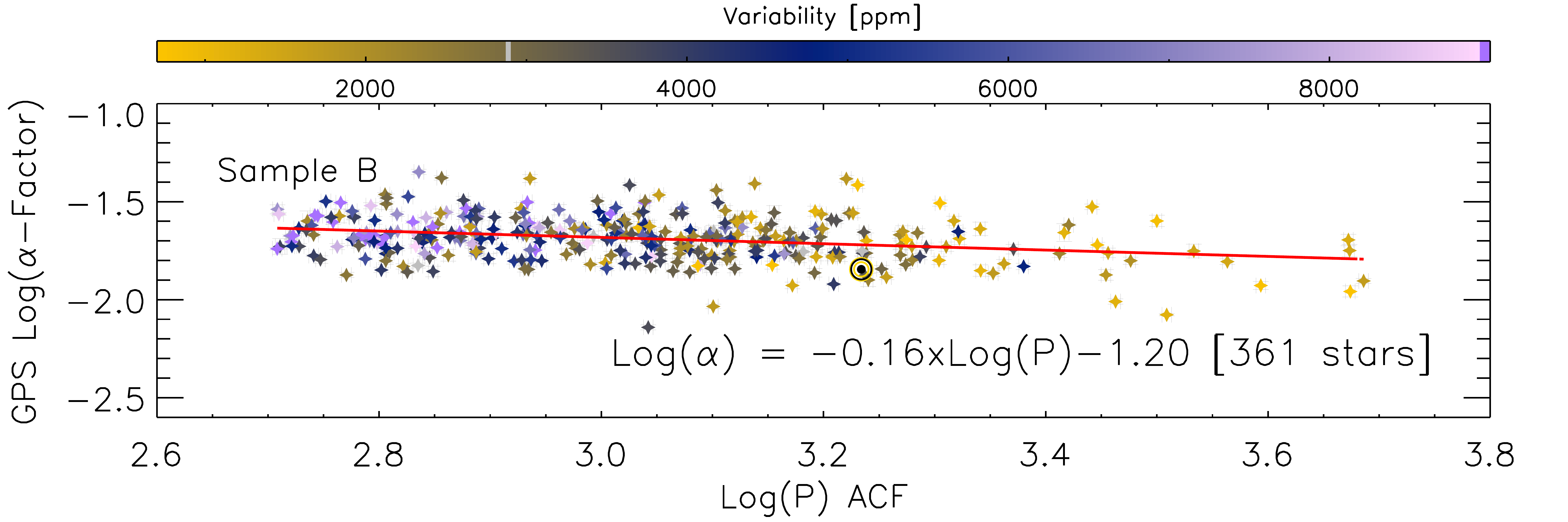}
\caption{Logarithmic visualization of bottom panel in Figure~\ref{fig4_5}.}
\label{fig4_5_Log}
\end{figure*}

\begin{figure*}[!ht]
\captionsetup{width=1.\linewidth}
\includegraphics[trim={0 0 0 0 cm},clip,width=1.\textwidth]{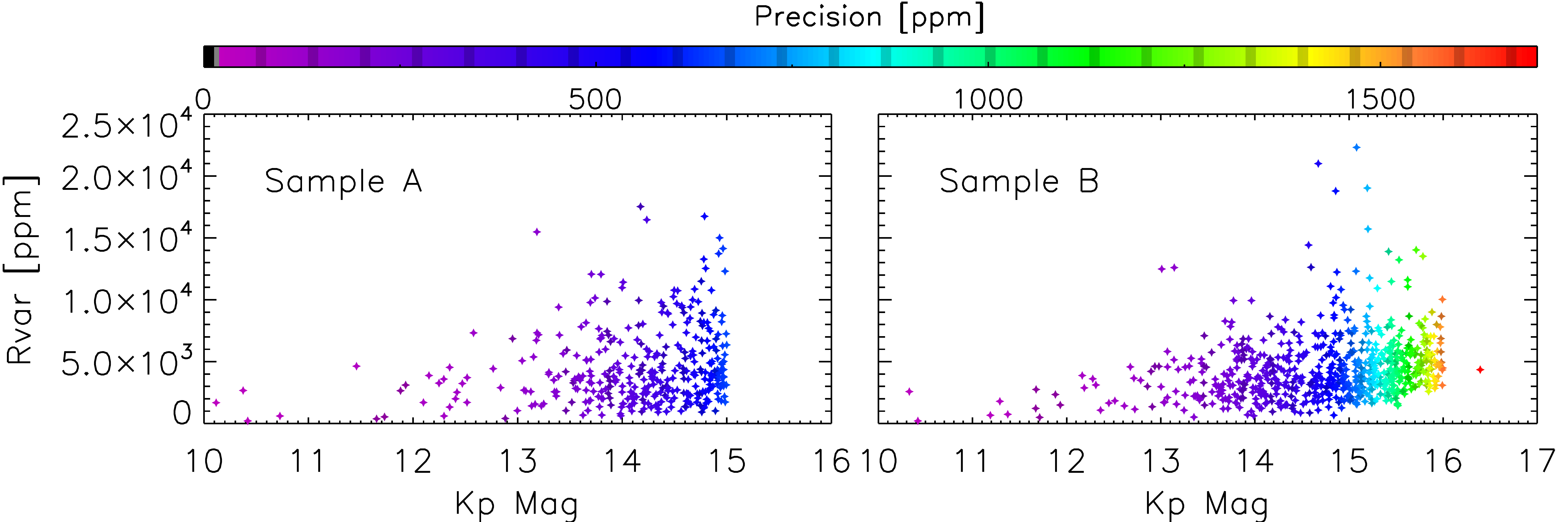}
\centering
\caption{Variability range in ppm versus Kepler magnitude Kmag. The colour bar indicates Kepler precision.}\label{RVAR_KEPMAG}
\end{figure*}

\end{appendix}

\end{document}